# Energy Allocation Policy for Cellular Networks Powered by Renewable Energy

**Qiao Li [1], Yifei Wei [1], Mei Song [1], F. Richard Yu[2]**

[1] School of Electronic Engineering, Beijing University of Posts and Telecommunications
Beijing, P.R. China
[e-mail: liqiao1989, weiyifei, songm@bupt.edu.cn]
[2] Department of Systems and Computer Engineering, Carleton University
Ottawa, ON, Canada
[e-mail: Richard.Yu@carleton.ca]
*Corresponding author: Qiao Li

## *Abstract*

The explosive wireless data service requirement accompanied with carbon dioxide emission and consumption of traditional energy has put pressure on both industry and academia. Wireless networks powered with the uneven and intermittent generated renewable energy have been widely researched and lead to a new research paradigm called green communication. In this paper, we comprehensively consider the total generated renewable energy, QoS requirement and channel quality, then propose a utility based renewable energy allocation policy. The utility here means the satisfaction degree of users with a certain amount allocated renewable energy. The energy allocation problem is formulated as a constraint optimization problem and a heuristic algorithm with low complexity is derived to solve the raised problem. Numerical results show that the renewable energy allocation policy is applicable for any situation. When the renewable energy is very scarce, only users with good channel quality can achieve allocated energy.

# 1. Introduction

**T**he continuously growing demand for ubiquitous wireless network access and high data rate lead to the rapid development of wireless cellular networks. The energy consumed to power the bigger and more complex networks, on one hand brings network operator the rapidly rising energy price, on the other hand, generates a large amount of carbon dioxide which results in climate change [1]. Green radio becomes an inevitable trend [2], and is implemented from two different perspectives. The first is to study novel techniques to improve the energy efficiency of wireless networks which can be measured by bits-per-Joule metric [3]-[4]. However for a fixed network size, this kind of energy efficiency design from the physical layer to the MAC layer has already approached the theory limits [5]. The second is to substitute renewable energy such as solar and wind power for traditional energy. The emerging trend of renewable energy powered wireless networks equipped with energy harvesting devices [6] has been widely studied recently. Wireless energy harvesting has been considered as a potential technology for future 5G networks and has been intensively researched [7]. However, the uneven and intermittent intrinsic characteristics of the renewable energy make it a finite resource in renewable energy powered wireless network. Hence, how to efficiently use the renewable energy distributed in energy harvesting wireless network has always been an open issue and many valuable research results have been achieved up to now.

In [8], taking into account channel conditions and energy sources that are time varying, so as to maximize the throughput, an optimal energy allocation with energy harvesting constraints is proposed in several time slots and solved via the use of dynamic programming and convex optimization techniques. In [9]-[10], and the references therein, multi-terminal models, and energy harvesting transmitters and receivers are subsequently studied. Green energy optimization problem in cellular networks powered by hybrid energy which include green energy and traditional energy is studied in [11]-[12] to reduce the power consumption of traditional energy. The study of energy cooperation between base stations (BSs) has been widely researched owing to the recent advancement in smart grid [13] and wireless power transfer [14] technologies. Through a two-way energy transmission between BSs, the throughput-maximization oriented energy cooperation strategies are studied in [15]-[16]. To the best of our knowledge, all the previous work for renewable energy utilization does not consider the various type of quality of service (QoS) of diverse traffic type in the wireless network powered by renewable energy.

In this paper, comprehensively considering the QoS, channel quality and the total available renewable energy, we propose the utility based energy allocation algorithm in the renewable energy powered orthogonal frequency division multiple access (OFDMA) cellular network. The utility here refers to a function which describes the degree of user satisfaction with a certain amount of allocated energy. In every time slot, each transmitter is constrained to use at most the amount of stored energy currently available, although more energy may become available in the future slots. Considering the traffic type with the soft QoS which means the traffic requires a certain preferred resource but will still tolerate resource below this preferred value, the channel quality of each user and the total amount of renewable energy at a BS, we allocate the renewable energy available at each BS to its users in each time slot at the method through which maximize the total utility of all users at the BS. The utility based renewable energy allocation method can balance the fairness and efficiency at various conditions of the reserved renewable energy at each current timeslot, the channel qualities and the traffic type with specific QoS requirement. The contributions of this paper are as follows:

1) Firstly, compared to the existing energy utilization policies which aim at maximizing the throughput or green energy utilization rate, in this paper we allocated the finite amount of renewable energy at each BS from the QoS perspective, since the guarantee of QoS is the most important thing the network should do.
2) Secondly, through the utility based renewable energy allocation algorithm, the finite amount of renewable energy at each BS can be used with a maximum total utility.
3) Finally, although our energy allocation algorithm is proposed for soft QoS traffic type it is also suitable for hard QoS or best effort QoS traffic type when utility function is very steep or flat.

The remainder of the paper is organized as follows. In Section II, we describe the system model. The renewable energy powered BS with OFDMA system is considered in this section. Section III formulates the problem. Considering the utility function of soft QoS traffic type then we formulate the renewable energy allocation problem as the total utility maximization problem. In Section IV, we deduce the solution for the problem posed in Section III. Numerical results and discussions are stated in detail in Section V. We conclude this paper in Section VI and simultaneously present our future work.

## 2. System Model

In this section we first describe the system model and then analyze the energy demand of various type of traffic.

### 2.1 System Scenario

Consider a single cell downlink OFDMA network with renewable energy powered base station and N active users as shown in **Fig. 1**. The renewable energy sources can be solar and wind power. The total bandwidth, $W_{total}$ is divided into $N$ subchannels, each subchannel with a bandwidth of $W_{sub} = W_{total} / N$ for each user. We assume that subcarriers are allocated centrally and each subchannel cannot be assigned to more than one user to avoid interference among different users. At each transmission period, the transmit power is denoted as $p_k$. Assuming perfect channel state information in both transmitter and receiver, the maximum achievable data rate of the $k$th user, denoted as d$_k$, is

$$d_k = W_{sub} \log_2(1 + \frac{p_k |h_k|^2}{N_0 W_{sub}}) \tag{1}$$

where $N_0$ is the single-sided noise spectral density, and the channel frequency response of the $k$th user whose data is transmitted on the $k$th subchannel is denoted as $h_k$. Consequently, at each transmission period the transmit power for the $k$th user is as follows.

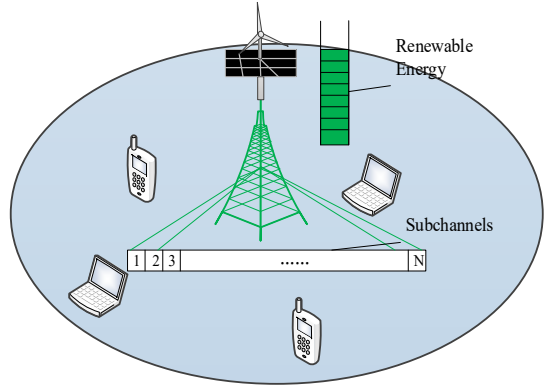

**Fig.1.** System Scenario

$$p_k = \frac{N_0 W_{sub}(2^{\frac{d_k}{W_{sub}}} - 1)}{|h_k|^2} \tag{2}$$

The total energy consumption of the BS during time slot $[t, t + \tau]$ is

$$E_k = \frac{N_0 W_{sub}(2^{\frac{d_k}{W_{sub}}} - 1)}{|h_k|^2} \tau. \tag{3}$$

For each user at each subchannel, there is an exponentially increasing relationship between the consumed energy $E_k$ and data rate $d_k$. The total energy consumed for all N users at N subchannels is

$$E_N = \sum_{k=1}^{N} \frac{N_0 W_{sub}(2^{\frac{d_k}{W_{sub}}} - 1)}{|h_k|^2} \tau. \tag{4}$$

### 2.2 Energy Demand and Renewable Energy Supply

The equipment parameters of an outdoor 4G BS (which is generally called eNode B and widely deployed throughout the world) produced by FUJITSU [17] are shown in **Table 1**. The maximum transmit power of eNodeB is 60 W and the transmit power will fluctuate with the traffic load, channel quality and QoS requirement of users.

We consider solar panels as the green energy generators. Solar panels generate electrical power by converting solar radiation into direct current electricity using semiconductors that exhibit the photo-voltaic effect, more information about the green energy generation and rate can be found in [18] and its reference therein [19]-[20]. We conduct an experiment that measures the power of the solar panel with the model SYK20-18P ( size: 540*350*25mm ) made by Guangzhou SUMYOK. We observed that the maximum power is 20W and solar energy generation depends on various factors, such as the temperature, the solar intensity, and the geolocation of the solar panels. Taking into account of the deployment cost of solar panel (or, windmill generator) and green energy storage devices and the time varying green energy generator rate, the renewable energy generating devices at each BS cannot do infinitely great and the renewable energy currently generated at each BS does not always exceed the transmit power of BS. These factors make renewable energy at each BS a finite resource.

**Table 1**. eNode B equipment specifications

| Item | Specifications |
|---|---|
| Radio frequency band | Band4, Band9, Band17 |
| Bandwidth | 5,10,15,20 MHZ |
| Access scheme | Downlink: OFDMA<br>Uplink: SC-FDMA |
| Maximum transmit power | 60 W (30W+30W) |
| Maximum transmission rate (per sector) | Downlink: 150Mb/s<br>Uplink: 50Mb/s |

Markov models have been widely accepted in the literature [21]-[25] as an effective approach to characterize the correlation structure of the fading process,Hence, we consider allocate the finite renewable energy based on the demand of diverse traffic type from various application. Assume that the bandwidth allocated to each subchannel is constant, and from Formula (1) the data rate of each user is mainly determined by the allocated transmit power from BS and the channel quality. Set the data rate as the main QoS metric, and a certain traffic type in the wireless network needs a specific QoS. Generally, under the current network structure, there are three kinds of QoS implementation models. The best effort QoS makes best effort to transfer data packet, but provides no guarantees and no priorization for users. Original internet service such as e-mail, file transfer and remote login belong to this type. The hard QoS has an explicit reservation of network resources for traffic flows before communication starts with strict resource requirement and poor scalability. Applications such as video conference, audio/video phone and tele-medicine need a hard QoS guarantee. As an intermediate step between hard QoS constraints and a pure best effort approach, soft QoS has a flexibility in network resource supply. The soft QoS traffic usually has intrinsic resource requirements, i.e., they have their own preferable resource values, but can still tolerate with a less amount of resource than the preferred amount. Soft QoS traffic also known as elastic traffic, can gracefully adjust their transmission rates to adapt to various network conditions. Interactive multimedia services, video on demand and most applications in current wireless networks are typical examples of soft QoS which has a high adaptability to network conditions and resources.

## 3. Problem Formulation

In this section we first define the utility function then formulate our renewable energy allocation problem. How to model application adaptation for utility function, and what type of overall system utility maximisation can be employed, are stated orderly in detail.

### 3.1 Utility Function

*Preference and Utility*: Preference relations are a handy way of talking about how people rank bundles of goods. We mainly use three binary relations to talk about preferences: $\succ$ (strictly preferred to), $\sqcup$ (indifferent between), and $\succeq$ (weakly preferred to). Utility function is used to easily describe preferences. A utility function assigns numerical values to all bundles so that if $x \succeq y$, we have $u(x) \geq u(y)$. As shown in Formula (1), set bandwidth $W_{sub}$ and the parameters $N_0$ and $h_k$ which represent the channel quality as constant, and assume the data rate as the main utility performance metric. Then the utility value is determined by the transmit power $p_k$. If there are two values $p_1$, $p_2$ of $p_k$ and $p_1 \geq p_2$, i.e., $p_1 \succeq p_2$, then $u(p_1) \geq u(p_2)$.

*Utility Application*: Utility function have been applied in wireline and wireless network resource allocation for many years. The utility function based resource allocation was first proposed in [26] and recently continued to be researched in [27] etc. mainly conclude the utility of time, spectrum, and wireless radio resources. The existing works have augmented the utility function model by identifying and classifying the way allocations affect the utility of different application classes. Generally, the utility based resource allocation problems usually have an optimal object which subjects to some network constraint.

*Utility Function*: In this paper, we studied the utility of energy, since the renewable energy generated at each BS is a finite and dynamic resource. As analyzed above, there is an exponential increasing relationship between utility and the renewable energy. A precise presentation of the utility function of renewable energy should comprehensively take into account of the total available energy, the channel quality and the traffic type. Consider that most applications in current networks are resource-adaptive and the network can allocate energy to users at a flexible way, so that we give attention to the utility function for the soft QoS traffic type. Mathematically, the following function can be used to model the utility function of soft QoS traffic type [28]-[29],

$$U(r) = \begin{cases} q \cdot e^{p(r-rmid)}, (r < rmid) \\ 1 - (1-q) \cdot e^{-p(r-rmid)}, (otherwise) \end{cases} \quad (5)$$

where *r* is the renewable energy allocated by the BS through a centralized mode, i.e., from the BS to users with diverse traffic flow at different subchannels, $q$ ($0 \leq q \leq 1$) denotes the channel quality of the user, *p* determines the slope of the utility function, and *rmid* denotes the preferable amount of resource for the soft QoS traffic. The marginal utility function is defined by the following equation, $u(r) = \dfrac{dU(r)}{dr}$, which is the first derivative of the utility function U(r) with respect to the given resource *r*. As shown in **Fig. 2**, when the allocated renewable energy *r* equals *rmid*, the marginal utility function achieve its maximum value.

Suppose that all traffic flows in a BS system have the same traffic type with the same QoS requirement. According to the utility function stated above, the amount of renewable energy allocated to each specific user at a BS determined by the channel quality q and traffic type of the user to guarantee the level of satisfaction. From the perspective of the whole BS system with finite renewable energy resource we should consider the amount of available renewable energy stored in the last inventory process step and channel conditions of all users to maximize the total utility.

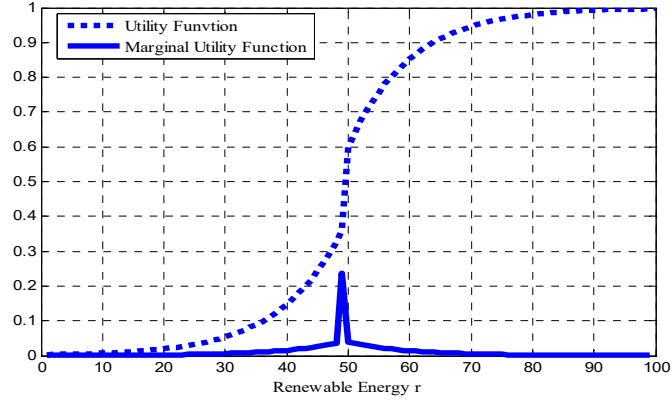

**Fig.2.** Utility function and marginal utility function

### 3.2 Problem Formulation

Assume there are $N$ users at our BS system. The total amount of available renewable energy at current energy allocation period is $r_{tot}$, and the energy allocated to the $k$th user is denoted as $r_k$. The traffic type and channel quality of each user may not be identical, so that the utility function and levels of satisfaction with the same amount of renewable energy of each user are different. Considering an amount of renewable energy allocated to the $k$th user with the channel quality $q_k$, due to the path loss, the renewable energy actually available to the $k$th user is given by $E_k = q_k \cdot r_k$. The utility function of the $k$th user can be expressed as $U_k(r_k) = U(r_k \cdot q_k)$. Considering the channel condition of each traffic flow from corresponding user may not be identical, we denote the preferable amount of resource for the $k$th traffic flow by $rmid_k$. For the $k$th traffic flow, $k \in \{1, 2, \cdots, N\}$, $U_k(rmid_k) = U(rmid) = U_{mid}$. Thus the preferable amount of resource for the $k$th traffic flow $rmid_k = \dfrac{rmid}{q_k}$.

In this paper, we aim at maximize the total utility of all users at a BS system with a limited amount of renewable energy generated in the current energy allocation period. The problem is formulated as,

$$\max \sum_{k=1}^{N} U_k(r_k)$$

$$sub. \begin{cases} \sum_{k=1}^{N} r_k \leq r_{tot} \\ \forall r_k \geq 0 \end{cases} \quad (6)$$

$$(k \in \{1, 2, \cdots, N\})$$

where $r_{tot}$ is the total amount of renewable energy that can be allocated in this current energy allocation period. The renewable energy allocation $R^* = \{r_1, r_2, \cdots, r_N\}$ for $N$ users is referred as an optimal allocation if $R^*$ can make Formula (6) established, i.e., for all feasible allocation $R' = \{r_1', r_2', \cdots, r_N'\}$, $U(R^*) \geq U(R')$, where $U(R^*) = \sum_{k=1}^{n} U_k(r_k)$ and $U(R') = \sum_{k=1}^{n} U_k(r_k')$. The optimal allocation $R^*$ may not be unique here.

## 4. Solution

To efficiently solve the constrained maximization problem, we appropriately relax the constraint conditions then deduce a heuristic algorithm.

To keep allocation optimal in the wireless environment changing both in the traffic and energy domain, we need to periodically (re)allocate resources. The available finite energy stored at a BS in this current time period will be allocated to $N$ users at this BS by an optimal allocation algorithm through which makes the total utility value maximum. Due to that an optimal solution to this problem presented in Formula (6) is very hard to find and is dependent on the channel qualities and utility functions of traffic flows, we first relax the constraint conditions then adopt a heuristic renewable energy allocation algorithm for soft QoS traffic flow. According to the number of users being allocated energy, all energy allocation (EA) method can be classified into $EA_1$, $EA_2$, $\cdots EA_N$. Assume that traffic flows from different users at respective subchannels have the identical traffic type and different channel quality.

*Property 1:* In an optimal renewable energy allocation method, each allocated user $i$, i.e., $r_i > 0$, must have an identical marginal utility value $u_i(r_i)$.

*Proof:* Assume there exists an optimal allocation $R$ in which there are users $i$ and $j$, ($i \neq j$), whose allocated resources $r_i$ and $r_j$ satisfy $u_i(r_i) \neq u_j(r_j)$ and $u_i(r_i) > u_j(r_j) > 0$. There must exist a finite small value $\Delta r$, and another allocation $R'$ with $r_i' = r_i + \Delta r$ and $r_j' = r_j - \Delta r$, ($r_i' + r_j' = r_i + r_j$).

$$U_i(r_i') + U_j(r_j') = U_i(r_i + \Delta r) + U_j(r_j - \Delta r)$$

Put Formula (5) into above equation, and we get,

$$U_i(r_i^{'}) + U_j(r_j^{'}) = U_i(r_i) + U_j(r_j) + (u_i(r_i) - u_j(r_j))\Delta r > U_i(r_i) + U_j(r_j)$$

The derivation declares that allocation $R$ is not optimal which violates the assumption at the beginning.

*Property 2:* At most one user, say $i$, with $u_i^{'}(r_i) > 0$ can be allocated energy in an optimal renewable energy allocation.

*Proof:* Assume there exists an optimal renewable energy allocation $R$ which have more than one allocated user I with $u_i^{'}(r_i) > 0$ denoted as $u_i^{'}(r_i) > 0$ and $u_j^{'}(r_j) > 0$. There must exist another allocation $R^{'}$ which allocates users $i$ and $j$ respectively with $r_i^{'}$ and $r_j^{'}$, where $r_i^{'} = r_i + \Delta r$ and $r_j^{'} = r_j - \Delta r$, i.e., $r_i^{'} + r_j^{'} = r_i + r_j$.

$$\begin{aligned}
&U(R^{'}) - U(R) \\
&= (U_i(r_i + \Delta r) - U_j(r_j - \Delta r)) - (U_i(r_i) - U_j(r_j)) \\
&= U_i(r_i + \Delta r) - U_i(r_i) + U_j(r_j - \Delta r) - U_j(r_j) \\
&= \int_{r_i}^{r_i + \Delta r} u_i(r)dr - \int_{r_j - \Delta r}^{r_j} u_j(r)dr \\
&\cong \frac{u_i(r_i) + (u_i(r_i) + u_i^{'}(r_i))}{2}\Delta r - \frac{u_j(r_j) + (u_j(r_j) + u_j^{'}(r_j))}{2}\Delta r \\
&= \left(\frac{(u_i^{'}(r_i) + u_j^{'}(r_j))}{2}\right)\Delta r \\
&> 0
\end{aligned}$$

The derivation declares that $U(R^{'}) > U(R)$, which vilate the assumption that $R$ is an optimal renewable energy allocation.

According to the above two properties, to reduce the computational complexity and guarantee that the allocation performance is still tightly bounded to the optimal solution, we assume that each user $i$ which has been allocated energy has $u_i^{'}(r_i) \leq 0$, i.e., the amount of allocated energy $r_i$ must exceed its preferable amount $rmid_i$. We conclude that an optimal renewable energy allocation must satisfy the following three conditions:
1) All renewable energy reserved in last inventory process must be allocated.
2) All allocated users must have an identical marginal utility.
3) All allocated users have $u_i^{'}(r_i) \leq 0$.

Set $r$, *rmid*, and $p$ in utility function as constant and the value of the utility function is monotone increasing along with $q$. That is to say, with a same amount of renewable energy, the better the channel quality is, the larger is the utility value. Renewable energy allocation $R_K$={$r_1$, $r_2$, $\cdots$, $r_K$}, which allocates renewable energy to the first $K$ users sorted by their channel qualities is the one with the highest utility among $EA_K$. We can find $R_K$ through 4 steps.
1) First, for $k = \{1, 2, \cdots, K\}$, we decompose the marginal sigmoid utility function $u_k(r_k)$ into two parts and redefine the part of $r_k > rmid_k$, i.e., $u_k^{'}(r_k) < 0$, as a new decreasing function $\bar{u}_k(\cdot)$, where $\bar{u}_k(r) = u_k(r + rmid_k)$.

2) Second, accumulate the inverse function of $\bar{u}_k(\cdot)$ which is denoted as $\bar{u}_k^{-1}(\cdot)$, of all allocated users, denoted as $\bar{u}_{sum}^{-1}(\cdot)$, where $\bar{u}_{sum}^{-1}(\cdot) = \sum_{k=1}^{K} \bar{u}_k^{-1}(\cdot)$ then we get the aggregated marginal utility function $\bar{u}_{sum}(\cdot)$ by inversing $\bar{u}_{sum}^{-1}(\cdot)$.

3) Thirdly, calculate the residual renewable energy res which is the amount of renewable energy allocated to $\bar{u}_k(\cdot)$, then the aggregated utility $u_a = \bar{u}_{sum}(res)$.

4) Finally, based on $u_a$ we can obtain the amount of energy allocated to each user $k$ in allocation $R_k=\{r_1, r_2, \cdots, r_k\}$, which is given by $r_k = rmid_k + \bar{u}_k^{-1}(u_a)$, for $k = \{1, 2, \cdots, K\}$ and $r_k = 0$, otherwise. Based on $R_k$, we calculate $\Delta U_k = U(R_k) - U(R_{k-1})$. If $\Delta U_k < 0$, the optimal renewable energy allocation $R_{k-1}$ is achieved, otherwise return $R_k$. The algorithm program is illustrated in detail in **Table 2**.

Table 2. Algorithm programme

**Heuristic algorithm** Finding optimal renewable energy allocation

**Initialization:** R= {0, 0, …, 0}, U(R₀)=0, r_tot, k=0.
1: **for** $k = \{1, 2, \cdots, K\}$
  $\bar{u}_k(r) = u_k(r + rmid_k)$
  $\bar{u}_k^{-1}(\cdot)$=inverse ($\bar{u}_k(\cdot)$)
 **end**
2: **for** k={0, 1, …, K}
  $\bar{u}_{sum}^{-1}(\cdot) = \sum_{k=1}^{K} \bar{u}_k^{-1}(\cdot)$
  $\bar{u}_{sum}(\cdot)$=inverse ($\bar{u}_{sum}^{-1}(\cdot)$)
  $res = r_{tot} - \sum_k rmid_k$
  $u_a = \bar{u}_{sum}(res)$
  $r_k = rmid_k + \bar{u}_k^{-1}(u_a)$
  $R_k$ is achieved.
  $\Delta U_k = U(R_k) - U(R_{k-1})$
   **if** $\Delta U_k < 0$
    The optimal renewable energy allocation: $R_{k-1}$ is achieved.
   **else** return $R_k$.
  k=k+1
 **end**

## 5. Numerical results and discussions

First, we change the slop of utility function through changing the value of *p* which is set 0.1, 0.8, 3.2, 14 respectively, and observe how the parameter *p* impacts the utility function. As shown in **Fig. 3**, when the value of *p* changes from 0.1 to 14, the utility function changed from a very flat curve to a steep one. Then we set 20 traffic flows at 20 independent subchannels at a BS, and the channel quality of each subchannel *q* is randomly selected in the range [0, 1], from 0 to 1. The larger value of *q* means the better channel quality and $q = 1$ represents the best channel quality. We set the preferable amount of renewable energy *rmid* for the best channel quality $q = 1$ as 10, and set $r_{tot}$ i.e., the total available renewable energy at each current period to be allocated, as 100, 400, and 1200, respectively representing a scarce, moderate, and sufficient amount of renewable energy generated at the current allocation period.

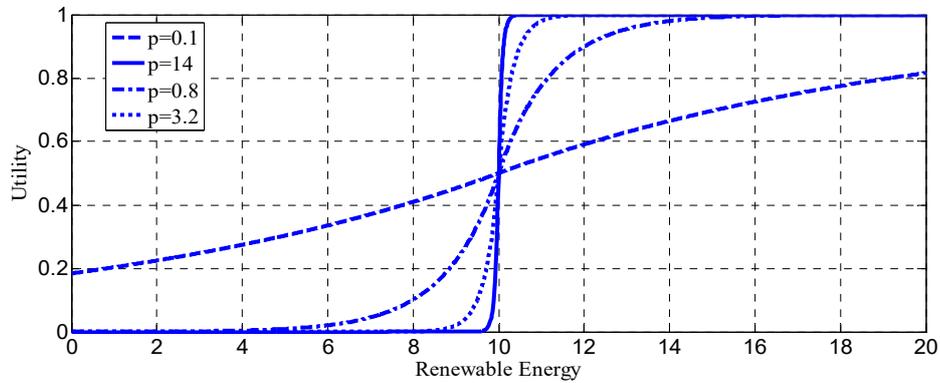

**Fig.3.** Utility function with different slope *p*.

Set the slope of utility function as 0.1. As shown in **Fig. 3**, the utility function is very flat. Set the amount of renewable energy avaliable at the current allocation period as 400, that is the BS has a moderate amount of renewable energy. As shown in **Fig. 4**, the better channel quality a user has, i.e., with a larger value of *q*, the more energy is actually available to the user. This illustrates the system allocates renewable energy to users in a way through which a maximum system throughput is achieved. In this case the energy allocation of the system acts in a best effort QoS method.

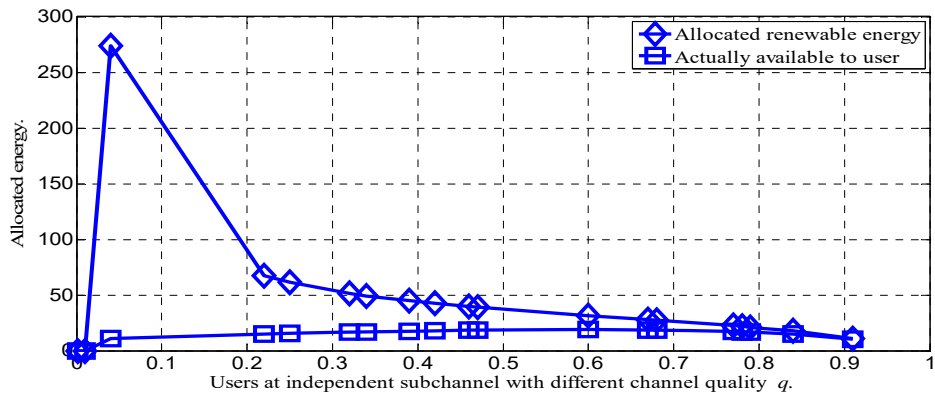

**Fig.4.** Allocated energy and the energy actually available to users: p=0.1, $r_{tot}$=400.

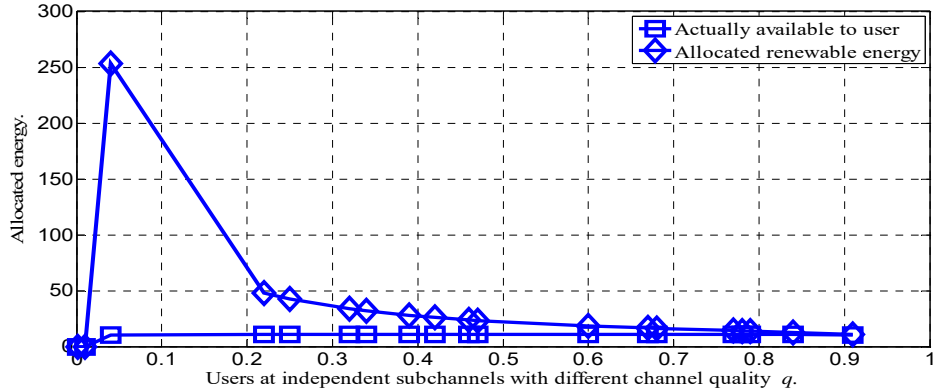

**Fig.5.** Allocated energy and the energy actually available to users: p=0.8, $r_{tot}$=400.

Set the slope of utility function as 14. As shown in **Fig. 3**, the utility function is very steep. Set the amount of renewable energy avaliable at the current allocation period as a moderate level, i.e., $r_{tot}$=400. **Fig. 6** shows that the system allocates more renewable energy to users with poorer channel quality, i.e., a smaller value of $q$, and the amount of energy actually available at each user in the system is the same. That is to say the system allocates energy in a fair way, i.e., no matter what channel quality the users have, the energy actually available to each user is the same. In this case the energy allocation of the system acts as a hard QoS method.

Set the total renewable energy generated at the current allocation period at a sufficient level, $r_{tot}$=1200. As shown in **Fig. 7**, no matter the slope of the utility function is flat or steep, i.e. p=0.1, 0.8, or 14, all users will be allocated a certain amount of renewable energy. The worse channel quality a user has, the more renewable energy will be allocated to the user. Set the total renewable energy generated at the current allocation period at a scarce level, $r_{tot}$=100. As shown in **Fig. 8**, no matter what kind of utility function is adopted, only users with good channel quality will be allocated renewable. It suggests that to achieve the goal of total utility maximization, the optimal renewable energy allocation is determined not only by the traffic type and channel quality, but also by the total amount of the available energy generated at the current allocation period.

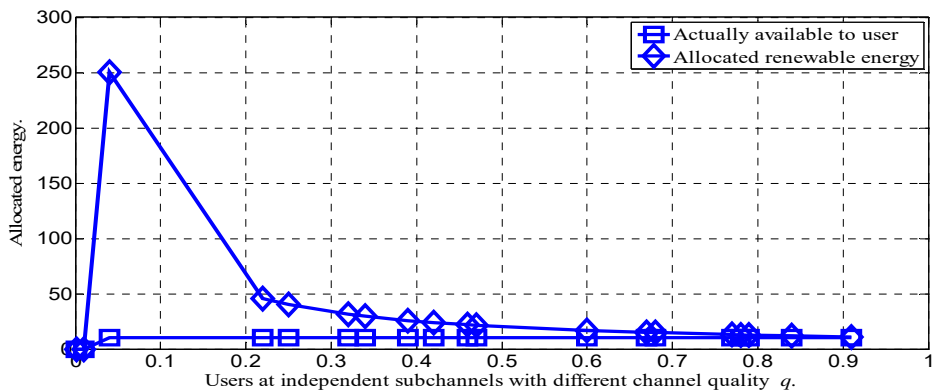

**Fig.6.** Allocated energy and the energy actually available to users: p=14, $r_{tot}$=400.

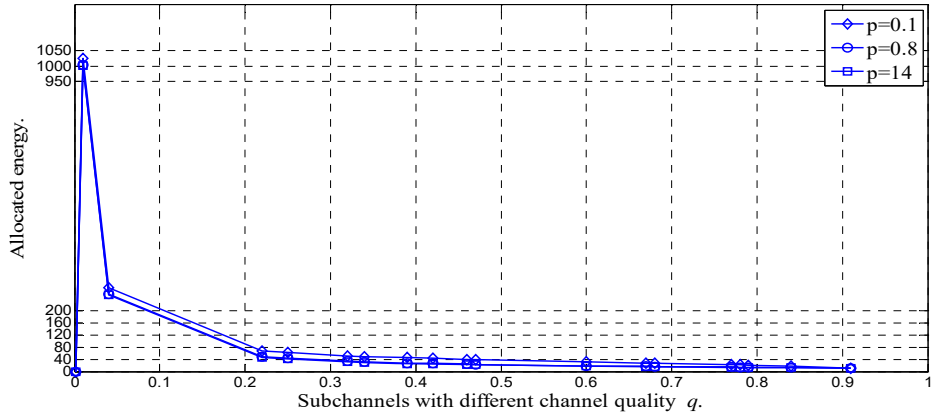

**Fig.7.** A sufficient amount of renewable energy, $r_{tot}=1200$.

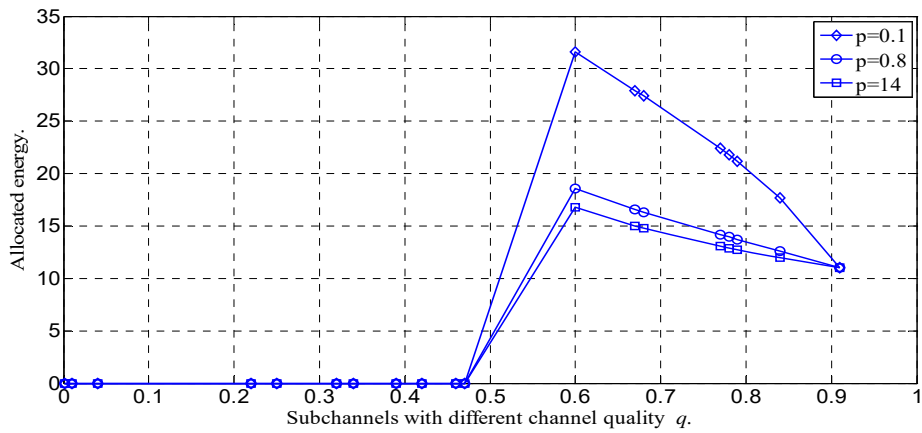

**Fig.8.** A scarce amount of renewable energy, $r_{tot}=100$.

## 6. Conclusion and future works

In this paper, we studied the renewable energy allocation policy which maximizes the total utility of all users at a BS with a finite amount of renewable energy. In our analysis, we comprehensively considered the traffic type, channel quality and total available renewable energy generated at each current allocation period. We took into account the soft QoS traffic type which most applications in the current wireless networks belongs to. Numerical results showed that when the total amount of renewable energy is moderate and the utility function is very steep, users at independent subchannels with various channel quality achieved the same amount of energy actually available to them. That is to say, when the utility function is pretty steep, the system provides a hard QoS and gives a fairness-oriented renewable energy allocation. When the total amount of renewable energy is moderate and the utility function is very flat, users only with a good channel quality can be allocated renewable energy. That is to say, when the utility function is pretty steep, the system provides a best effort QoS. While the total amount of renewable energy is scarce, no matter what kind of utility function we choose, the renewable energy will only be allocated to the users with good channel quality. However, in the renewable energy powered wireless networks there exists other important scarce resource, i.e., spectrum. In order to provide the users with a high quality of experience (QoE) we will jointly optimize the utilization of the finite renewable energy and spectrum in our future work.

## References

[1] A. Fehske, G. Fettweis, J. Malmodin, and G. Biczok, "The Global Footprint of Mobile Communications: The Ecological and Economic Perspective," *IEEE Communications Magazine*, vol. 49, no. 8, pp. 55-62, August, 2011.

[2] C. Han et al., "Green radio: radio techniques to enable energy-efficient wireless networks," *IEEE Communications Magazine*, vol. 49, no. 6, pp. 46-54, June, 2011.

[3] V. Rodoplu and T.H. Meng, "Bits-per-Joule Capacity of Energy Limited Wireless Networks," *IEEE Transactions on Wireless Communications*, vol. 6, no. 3, pp. 857-865, March, 2007.

[4] Nguyen Dinh Han, Yonghwa Chung, and Minho Jo, "Green Data Centers for Cloud-assisted Mobile Ad-hoc Networks in 5G," *IEEE Network*, vol.29, no. 2, pp. 70-76, April, 2015.

[5] Ritesh Kumar Madan, "Resource allocation algorithms for energy efficient wireless networks", Ph.D. dissertation, Stanford University, August, 2006.

[6] S. Ulukus, A. Yener, E. Erkip, O. Simeone, M. Zorzi, and K. Huang, "Energy harvesting wireless communications: A review of recent advances," *IEEE Journal on Selected Areas in Communications*, vol. 33, no. 3, pp. 360–381, March, 2015.

[7] Hongyuan Gao, Waleed Ejaz and Minho Jo, "Cooperative Wireless Energy Harvesting and Spectrum Sharing in 5G Networks," *IEEE Access*, vol.4, pp. 3647-3658, July, 2016.

[8] Chin Keong Ho and Rui Zhang, "Optimal Energy Allocation for Wireless Communications With Energy Harvesting Constraints", *IEEE Transactions on Signal Processing*, vol. 60, no. 9, pp. 4808 - 4818, September, 2012.

[9] I. Ahmed, A. Ikhlef, R. Schober, and R. K. Mallik, "Power allocation for conventional and buffer-aided link adaptive relaying systems with energy harvesting nodes," *IEEE Transactions on Wireless Communications*, vol. 13, no. 3, pp. 1182–1195, March, 2014.

[10] Y. Luo, J. Zhang and K. B. Letaief, "Optimal scheduling and power allocation for two-hop energy harvesting communication systems," *IEEE Transactions on Wireless Communications*, vol. 12, no. 9, pp. 4729-4741, September, 2013.

[11] H. Tao and N. Ansari, "On optimizing green energy utilization for cellular networks with hybrid energy supplies," *IEEE Transactions on Wireless Communications*, vol. 12, no. 8, pp. 3872–3882, August, 2013.

[12] Congshi Hu, Jie Gong, Xiaolei Wang, Sheng Zhou and Zhisheng Niu, "Optimal Green Energy Utilization in MIMO Systems With Hybrid Energy Supplies", *IEEE Transactions on Vehicular Technology*, vol. 64, no. 8, pp. 3675-3688, August, 2015.

[13] R. Ma, H.-H. Chen, Y.-R. Huang and W. Meng, "Smart grid communication: Its challenges and opportunities," *IEEE Transactions on Smart Grid*, vol. 4, no. 1, pp. 36–46, March, 2013.

[14] D. Kwan Ng., E. Lo and R. Schober, "Wireless information and power transfer: Energy efficiency optimization in OFDMA systems," *IEEE Transactions on Wireless Communications*, vol. 12, no. 12, pp. 6352–6370, December, 2013.

[15] Berk Gurakan, Omur Ozel, Jing Yang and Sennur Ulukus "Energy Cooperation in Energy Harvesting Communications" *IEEE Transactions on Communications*, vol. 61, no. 12, pp. 4884-4898, December, 2013.

[16] Jie Xu and Rui Zhang, "CoMP Meets Smart Grid: A New Communication and Energy Cooperation Paradigm" *IEEE Transactions on Vehicular Technology*, vol. 64, no. 6, pp. 2476 – 2488, June, 2015.

[17] Kimio Watanabe and Mamoru Machida, "Outdoor LTE Infrastructure Equipment (eNodeB)", *Fujitsu scientific & technical journal*, vol. 48, no. 1, January, 2012.

[18] T. Han and N. Ansari, "On optimizing green energy utilization for cellular networks with hybrid energy supplies," *IEEE Transactions on Wireless Communications*, vol. 12, no. 8, pp. 3872–3882, August, 2013.
[19] "System advisor model (SAM)." Available: https://sam.nrel.gov/
[20] "PVWatts." Available: http://www.nrel.gov/rredc/pvwatts/
[21] Yifei Wei, F. Richard Yu and Mei Song, "Distributed Optimal Relay Selection in Wireless Cooperative Networks with Finite State Markov Channels", IEEE Transactions on Vehicular Technology, vol. 59, no. 5, pp. 2149-2158, June 2010.
[22] L. Zhu and F. R. Yu and B. Ning and T. Tang, "Cross-Layer Handoff Design in MIMO-Enabled WLANs for Communication-Based Train Control Systems", IEEE J. Sel. Areas Commun., vol.30, no.4, pp. 719-728, 2012.
[23] Yifei Wei, Xiaojun Wang, LeonardoFialho, Roberto Bruschi, Olga Ormond, Martin Collier, "Hierarchical power management architecture and optimal local control policy for energy efficient networks", Journal of Communications and Networks, vol.18, no.4, pp.540-550, 2016.
[24] Zhao N, Yu F R, Leung V C M. Opportunistic communications in interference alignment networks with wireless power transfer[J]. IEEE Wireless Communications, 2015, 22(1): 88-95.
[25] Yifei Wei, ChenyingRen, Mei Song, Richard Yu,"The offloading model for green base stations in hybrid energy networks with multiple objectives", International Journal of Communication Systems, vol.29, no.11, pp.1805-1816, 2016.
[26] M. Sheng, C. Xu, X. Wang, Y. Zhang, W. Han and J. Li , "Utility-Based Resource Allocation for Multi-Channel Decentralized Networks," *IEEE Transactions on Communications*, vol. 62, no. 10, pp. 3610-3620, October, 2014.
[27] R. Deng, G. Liu and J. Yang, "Utility-Based Optimized Cross-Layer Scheme for Real-Time Video Transmission Over HSDPA", *IEEE Transactions on Multimedia*, vol. 17, no. 9, pp. 1495-1507, September, 2015.
[28] S. Shenker, "Fundamental Design Issues for the Future Internet," *IEEE Journal on Selected Areas in Communications*, vol. 13, no. 7, pp. 1176-1188, September, 1995.
[29] Liansheng Tan, Zhongxun Zhu, Fei Ge and Naixue Xiong, "Utility Maximization Resource Allocation in Wireless Networks: Methods and Algorithms", *IEEE Transactions on Systems, Man, and Cybernetics: Systems*, vol. 45, no.7, pp. 1018-1034, July, 2015.